\documentstyle[epsf,graphics,epsfig,times]{mn}

\def\x{\mbox{\boldmath $x$}}
\def\xp{\mbox{\boldmath $x'$}}
\def\C{\mbox{\boldmath $C$}}
\def\D{\mbox{\boldmath $D$}}
\def\be{\begin{equation}}
\def\ee{\end{equation}}
\def\bm{\begin{displaymath}}
\def\em{\end{displaymath}}
\def\hompc{\,h\,{\rm Mpc}^{-1}}
\def\mpcohcube{\,h^{-3}\,{\rm Mpc}^{3}}

\font\japit = cmti10 at 11truept

\title[2dFGRS: real-space power spectrum] {\vglue-3.0truecm \centerline{\japit
Submitted for publication in Monthly Notices of the R.A.S.}  \vglue
2.5truecm\noindent Markov-chain reconstruction of the 2dF Galaxy
Redshift Survey real-space power spectrum}

\date{Submitted for publication in MNRAS}
\begin{document}

\author[W.J.~Percival]{
\parbox[t]{\textwidth}{Will J.\ Percival}
\vspace*{6pt} \\ 
Institute for Astronomy, University of Edinburgh, Royal Observatory, 
Blackford Hill, Edinburgh EH9 3HJ, UK \\
}

\date{}

\maketitle

\begin{abstract}
The real-space power spectrum of $L_*$ galaxies measured from the 2dF
Galaxy Redshift Survey (2dFGRS) is presented. Markov-Chain Monte-Carlo
(MCMC) sampling was used to fit radial and angular modes resulting
from a Spherical Harmonics decomposition of the 2dFGRS overdensity
field (described in Percival et~al. 2004) with 16 real-space power
spectrum values and linear redshift-space distortion parameter
$\beta(L_*,0)$. The recovered marginalised band-powers are compared to
previous estimates of galaxy power spectra. Additionally, we provide a
simple model for the 17 dimensional likelihood hyper-surface in order
to allow the likelihood to be quickly estimated given a set of model
band-powers and $\beta(L_*,0)$. The likelihood surface is not well
approximated by a multi-variate Gaussian distribution with
model-independent covariances. Instead, a model is presented in which
the distribution of each band-power has a Gaussian distribution in a
combination of the band-power and its logarithm. The relative
contribution of each component was determined by fitting the MCMC
output. Using these distributions, we demonstrate how the likelihood
of a given cosmological model can be quickly and accurately estimated,
and use a simple set of models to compare estimated likelihoods with
likelihoods calculated using the full Spherical Harmonics procedure.
All of the data are made publically available (from {\tt
http://www.roe.ac.uk/{\tt\char'176}wjp/}), enabling the Spherical
Harmonics decomposition of the 2dFGRS of Percival et al. (2004) to be
easily used as a cosmological constraint.
\end{abstract}


\section{introduction}

In addition to a wealth of information about the physics of galaxy
formation, galaxy surveys also provide a statistical measure of the
large-scale primordial density perturbations. If, as is theoretically
expected, the density field on large-scales is well-approximated by a
random-phase Gaussian distribution, then the power spectrum provides a
statistically complete description of the field. Given a galaxy
redshift survey, the most direct route to an approximation to the
power spectrum is to decompose the overdensity distribution into
Fourier modes (e.g. Feldman, Kaiser \& Peacock 1994). However,
systematic observed distortions in the radial direction caused by the
coherent peculiar velocities of galaxies are difficult to incorporate
into a model of a Fourier decomposition.

Consequently, methods to decompose the galaxy overdensity field into
3-dimensional bases consisting of radial and angular functionals have
been developed by a number of groups (Heavens \& Taylor 1995; Fisher
et~al. 1994;1995; Hamilton, Tegmark \& Padmanabhan 2000). By splitting
the basis into radial and angular components, radial redshift-space
distortions can be readily included in the analysis
method. Recently, the 2dFGRS and SDSS galaxy redshift surveys have
both been analysed in this way (Percival et~al. 2004:P04, Tegmark et
~al. 2004) resulting in tight constraints on both the power spectrum
shape and the redshift-space distortions. Other analyses of 2-point
statistics measured from these surveys include Percival et~al. (2001)
and Cole et~al. (2004) who decomposed the 2dFGRS into Fourier modes
and Pope et~al. (2004) who undertook a Karhunen-Lo\`{e}ve eigenmode
analysis of the SDSS based on the correlation function of
counts-in-cells.

In this paper we build upon the analysis of P04 who decomposed the
2dFGRS overdensity field into an orthonormal basis of spherical
harmonics and spherical Bessel functions. Throughout we refer to such
a basis as a Spherical Harmonics basis. In P04, cosmological models
were tested against combinations of the radial and angular modes
calculated for the decomposed density field. The procedure for
calculating the likelihood of each model required inversion of a large
covariance matrix ($\sim1800\times1800$) and was computationally
expensive. It would obviously be advantageous if the same Likelihood
surface could be recovered without having to invert a large matrix for
each cosmological model.

The Markov-chain Monte-Carlo (MCMC) technique provides a method for
mapping a likelihood surface, providing both marginalised parameters
and information about the form of the surface. In this paper, we use
the MCMC technique to map the true likelihood surface as a function of
$16$ power spectrum band-powers and $\beta(L_*,0)$. Here
$\beta(L_*,0)\equiv\Omega_m^{0.6}/b(L_*,0)$ is a measure of the
increased fluctuation amplitude for present day $L_*$ galaxies caused
by the linear movement of matter onto density peaks and out from voids
(Kaiser 1987). $\Omega_m$ is the present day matter density and
$b(L_*,0)$ is the present day bias of $L_*$ galaxies. The resulting
sampling of the likelihood surface provides an estimate of the
real-space galaxy power spectrum through the marginalised power
spectrum band-powers (presented in Section~\ref{sec:real_pk}).  In
addition, we fit models to the distribution of each band-power,
marginalised over the other parameters
(Section~\ref{sec:like_fit}). The distributions are assumed to be
Gaussian in a combination of the band-power and its logarithm. Simply
assuming a Gaussian distribution in each parameter and
model-independent covariances could potentially bias the likelihood
estimates. This is directly analogous to the problem of compressing
CMB maps into angular band-powers, and the use of those band-powers to
estimate the true likelihood surface (e.g. Bond Jaffe \& Knox
2000). The model of the likelihood surface can be used to quickly
estimate the likelihood of any theoretical power spectrum and
$\beta(L_*,0)$. In Section~\ref{sec:simple_models} the likelihood
surface estimated for a simple set of cosmological models from these
distributions is shown to be in excellent agreement with the original
likelihood surface.

\section{the Spherical Harmonics decomposition}

This paper builds upon the Spherical Harmonics decomposition of the
2dFGRS overdensity field presented in, and completely described by
P04. This description is not replicated here. Instead, we discuss the
key points of the P04 analysis which are relevant for the results
presented in this paper. This includes a detailed discussion of the
likelihood calculation performed in P04 as this provides the benchmark
against which we test the accuracy that can be achieved from the
compressed data set.

\subsection{The methodology of P04}

The formalism used by P04 was based in part on that developed by
Heavens \& Taylor (1995) and described by Tadros et~al. (1999). First,
the galaxy overdensity field is decomposed into an orthonormal basis
consisting of spherical Bessel functions and spherical harmonics. A
constant galaxy clustering (CGC) model is assumed, where the amplitude
of comoving galaxy clustering is independent of redshift, although it
is assumed to be dependent on galaxy luminosity through the relation
of Norberg et~al. (2001). Each galaxy is weighted using this model so
that the measured power spectrum corresponds to that of $L_*$ galaxies
(using the method developed in Percival, Verde \& Peacock 2004).

To calculate the likelihood of a given cosmological model, the
expected covariances between modes have to be calculated. The primary
difficulty with this calculation is correctly accounting for the
geometry of the 2dFGRS sample, which results in a significant
convolution of the true power, and is performed as a discrete sum over
Spherical Harmonic modes in a computationally intensive part of the
adopted procedure. For computational convenience, and to simplify the
real-space shape of the window function, the 2dFGRS sample was split
into two wedge shaped components -- the NGP and SGP regions of the
2dFGRS (Colless et~al. 2003).

In addition to direct corrections for the survey geometry, the effects
of redshift-space distortions are also included at this stage. A first
order correction is made for the large-scale redshift-space
distortions which are linearly dependent on the density field through
$\beta(L_*,0)$. Because we decompose along radial modes, a far-field
approximation (Kaiser 1987) is not required, and the geometry of the
distortions is correctly accounted for. A contribution to the observed
radial clustering from the small-scale velocity dispersion of galaxies
is also included using an additional convolution in the radial
direction. The modes chosen cover sufficiently large scales that the
exact form of this convolution was unimportant. This convolution
provides a statistical model for the fingers-of-god distortions and
could be replaced by a more direct method such as the compression of
clumps along the line of sight (as favoured by Tegmark et~al. 2004).

The transformation between the Spherical Harmonics and the 3D Fourier
basis is unitary, so the length of a complex vector is
preserved. Consequently, for each Spherical Harmonics basis element we
can associate a Fourier $k$-mode, and a Spherical Harmonics
decomposition can be directly related to the standard Fourier power
spectrum. The P04 decomposition of the overdensity field resulted in
$86667$ Spherical Harmonic modes with $0.02<k<0.15\hompc$. This range
of $k$ was chosen to cover most of the cosmological signal without
excessive contamination from small-scale effects (see Percival
et~al. 2001 for details). It is impractical to directly use all of
these modes in a likelihood analysis, and the modes were therefore
compressed, leaving $1223$ \& $1785$ combinations of modes for the NGP
and SGP respectively. The data compression procedure adopted was
designed to remove nearly degenerate modes, which could cause
numerical problems followed by Karhunen-Lo\`{e}ve data compression to
optimally reduce the remaining data. Tests have shown that the
inclusion of further modes did not significantly improve the
cosmological results from the analysis.  The speed of the analysis
method depends on the cube of the number of modes: with the current
mode number, the analysis of P04 takes approximately 1 month to run on
a single top-of-the-line desktop computer, to perform the
decomposition and calculate the covariance matrix components. The MCMC
analysis of the likelihood surface presented in this paper takes
slightly longer. The compressed data and the corresponding covariance
matrices are combined to calculate the likelihood of a given model
assuming Gaussian statistics as described in the next Section.

\subsection{Likelihood calculation}  \label{sec:like_true}

For a Gaussian random field, the decomposition of the 2dFGRS
overdensity field obviously follows Gaussian statistics. The
likelihood of a cosmological model parameterized by the power spectrum
$P(k)$ and linear redshift-space distortion parameter $\beta(L_*,0)$
is given by
\bm
  {\cal {L}}_{\rm true}[\D|\beta(L_*,0),P(k)] = 
\em
\be
  \hspace{1cm}\frac{1}{(2 \pi)^{N/2} |{\C}|^{1/2}}
    {\rm exp}\left[ - \frac{1}{2} \D^T \C^{-1} \D\right],
    \label{eq:like_true}
\ee
where $N$ gives the number of data points to be tested, $C$ is the
covariance matrix, and $D$ is a linear combination of the Spherical
Harmonics modes. Note that the likelihood depends on the cosmological
model solely through the covariance matrix $C$ as $\langle
D\rangle=0$, independent of cosmology. For computational convenience
the covariance matrix was calculated for each model from a series of
components covering different ranges in $k$. Consequently, even for
the ``true'' likelihood as determined in P04, the data are effectively
compressed, although $30$ intervals in $k$ are used rather than
$16$. Because of this, the likelihood comparison presented in
Section~\ref{sec:simple_models} does not exhaustively test the data
compression part of the revised method. The likelihoods for the NGP
and SGP regions of the 2dFGRS were calculated separately, and were
then combined assuming statistical independence, which was shown to be
a valid approximation using mock catalogues.

As described in the previous section, the calculation of the
covariances related to the Spherical Harmonic modes requires a
convolution of the galaxy power spectrum, allowing for the geometry of
the survey and redshift-space distortions, and an additive shot noise
term. In the absence of distortions, for an all-sky volume-limited
survey with no noise, the modes are independent and their variance
reduces to the power spectrum at the $k$-value corresponding to that
Spherical Harmonic mode. This is analogous to analysis of all-sky
noiseless CMB data (e.g. Bond et~al. 2000; Verde et~al. 2003), and
results in a likelihood of the form
\begin{equation}
  -2\ln{\cal L} = \sum_{\rm modes} \left[ \ln P(k_i)^{\rm TH} 
    + \frac{\bar\delta^2(k_i)}{P(k_i)^{\rm TH}} \right],
  \label{eq:like_allsky}
\end{equation}
up to an irrelevant additive constant. Here $\bar\delta(k_i)$ gives
the observed transformed overdensity field, and $i$ indexes the
Spherical Harmonics modes.

\subsection{Choice of band-powers}  \label{sec:bands_choice}

In the analysis presented in this paper, the power spectrum was
assumed to be constant within 16 bins in $k$-space covering the range
of scales probed by the Spherical Harmonics analysis of P04. Although
only Spherical Harmonic modes corresponding to $0.02<k<0.15\hompc$
were analysed, the geometry of the survey means that these modes have
windows that extend beyond this range, and the power spectrum was
therefore probed on both larger and smaller scales. However, the power
spectrum is expected to become increasingly poorly constrained for
$k<0.02\hompc$ and $k>0.15\hompc$ simply because modes with windows
centred in these regions were excluded. The widths of the $k$-space
bins used (given in Table~\ref{tab:pk}) were chosen to reflect this.

Although each Spherical Harmonic mode has an associated window
function, band-powers were fitted to the underlying unconvolved power
spectrum. Because of this, the measured band-powers do not have
windows: they do not cover an extended range in $k$. Model band-powers
can therefore easily be calculated for a given cosmological model by
averaging the model power spectrum across each bin. In addition to
fitting the power spectrum shape through the band-powers,
$\beta(L_*,0)$ is also allowed to vary freely. The 17-dimensional
likelihood space for these parameters has been mapped using the MCMC
technique.

\section{MCMC method}  \label{sec:mcmc}

The Markov Chain Monte Carlo (MCMC) method provides a mechanism to
generate a random sequence of parameter values whose distribution
matches the posterior probability distribution of a Bayesian
analysis. This Section describes the method by which 4 chains, each
containing $>60000$ links, were produced mapping the 17-dimensional
likelihood space consisting of the 16 band powers and $\beta(L_*,0)$.

The 4 chains were started from widely separated points in parameter
space and were constructed from this point using the Metropolis
algorithm (Metropolis et~al. 1953): given a chain at position $\x$, a
candidate point $\xp$ is chosen at random from a proposal distribution
$f(\xp|\x)$. This point is always accepted, and the chain moves to
point $\xp$, if the new position has a higher likelihood. If the new
position $\xp$ is less likely than $\x$, then $\xp$ is accepted, and
the chain moves to point $\xp$ with probability given by the ratio of
the likelihood of $\xp$ and the likelihood of $\x$. In the limit of an
infinite number of steps, the chains will reach a converged
distribution where the distribution of chain links are representative
of the likelihood hyper-surface, given any symmetric proposal
distribution $f(\xp|\x)=f(\x|\xp)$ (the Ergodic theorem: see, for
example, Roberts 1996).

In the implementation of the MCMC method used in the 2dFGRS analysis
presented in this paper, the proposal distribution was altered as the
algorithm progressed in order to optimize the sampling of the
likelihood surface. Dynamic optimization of the proposal function is a
common feature of many MCMC analyses (see Gilks, Richardson \&
Spiegelhalter 1996 for examples). A multi-variate Gaussian proposal
function was adopted, centered on the current chain position. The
optimization of this proposal function requires an estimate of the
covariance matrix for the likelihood surface at each step. Given such
an estimate, the optimization proceeds as follows:

Along each principal direction corresponding to an eigenvector of the
covariance matrix, the variance $\sigma^2$ of the multi-variate
Gaussian proposal function was assumed to be a fixed multiple of the
corresponding eigenvalue of the covariance matrix. The reasoning
behind adopting this variation in the proposal width for different
parameters is clear if the likelihood also has a multi-variate
Gaussian form. Consider translating from the original 17 parameters to
the set of parameters given by the decomposition along the principal
directions of the covariance matrix each divided by the standard
deviation in that direction. In this basis, the likelihood function is
isotropic and the parameters are uncorrelated. Clearly an optimized
proposal function will be the same in each direction, and we have
adjusted the proposal function to have precisely this property. There
is just a single parameter left to optimize -- we are free to multiply
the width of the proposal function by a constant in all
directions. This overall normalization of the proposal width was
dynamically adjusted to optimize chain convergence by giving a final
acceptance rate, the proportion of candidate positions that are
accepted, close to $0.25$ (Gelman, Roberts \& Gilks 1996).  Note that
the dynamic changing of the proposal function width violates the
symmetry of the proposal distribution $f(\xp|\x)$ assumed in the
Metropolis algorithm. However, this was not a problem because we only
used sections of the chains where variations between estimates of the
covariance matrix were small.

Initially, a diagonal covariance matrix was assumed with a large
variance for each of the 17 parameters, in order to quickly cover
large regions of the 17-dimensional space. This estimate was then
updated every $100$ steps of the algorithm using the distribution of
parameters within the last half of all 4 chains. The covariance matrix
was found to rapidly converge to a stable solution. The dynamic
covariance matrix calculation means that the step size decreases from
a high initial value towards the value giving an average acceptance
ratio of $0.25$. The decreasing step-size reduces the burn-in time
where the position of the chain is strongly dependent on the initial
conditions. By considering the likelihood as a function of chain
position, we adopt a very conservative value for the burn-in, and
discard the first $10000$ links from each chain. Convergence was
confirmed using the method of Gelman \& Rubin (1992), also described
in Verde et~al. (2003), although we have additionally tested that
using only the first or second half of the 4 chains does not
significantly affect the results of the analysis.

\section{the 2\lowercase{d}FGRS real-space power spectrum}  \label{sec:real_pk}

\begin{figure}
  \setlength{\epsfxsize}{0.95\columnwidth} 
    \centerline{\epsfbox{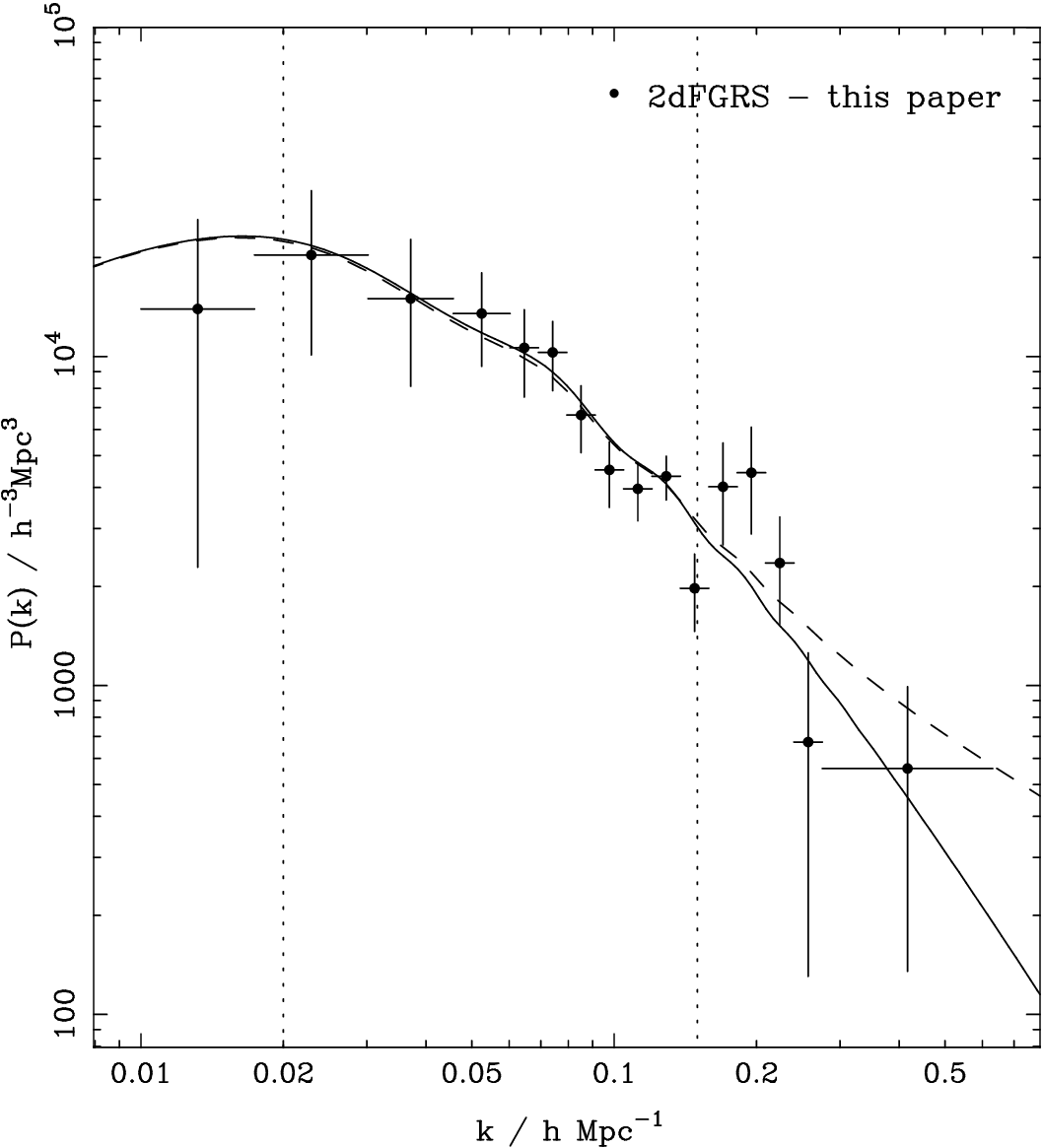}}

  \caption{The real-space power spectrum of $L_*$ galaxies calculated
  from the 2dFGRS (solid symbols). For comparison we also plot a model
  linear power spectrum with $\Omega_mh=0.2$,
  $\Omega_b/\Omega_m=0.15$, $b(L_*,0)\sigma_8=0.8$, $n_s=1.0$ and
  $h=0.72$ (solid line). The dashed line shows that non-linear
  extrapolation of this model, calculated using the fitting formulae
  of Smith et~al. (2003). Vertical dotted lines show the $k$-range
  chosen for the Spherical Harmonic modes fitted
  $0.02<k<0.15\hompc$. \label{fig:real_pk_2dF}}
\end{figure}

\begin{table}
  \caption{Marginalised real-space power spectrum band-powers
  calculated using the MCMC technique. Minimum and maximum
  $P(k)$ values correspond to 1-$\sigma$ errors and were calculated
  using the form adopted for the likelihood
  (Eq.~\ref{eq:like_special}).  \label{tab:pk}}

  \centering \begin{tabular}{ccccc} \hline
  \multicolumn{2}{c}{$k$ / $\hompc$} & 
  \multicolumn{3}{c}{$P(k)$ / $\mpcohcube$} \\ 
  min & max & marg & min & max \\
  \hline
  0.0100 & 0.0174 &   13944.9 &    2289.3 &   26070.9 \\ 
  0.0174 & 0.0302 &   20343.2 &   10112.6 &   31912.1 \\ 
  0.0302 & 0.0457 &   14999.7 &    8118.1 &   22685.4 \\ 
  0.0457 & 0.0603 &   13505.5 &    9336.3 &   17958.2 \\ 
  0.0603 & 0.0692 &   10629.0 &    7528.0 &   13870.8 \\ 
  0.0692 & 0.0794 &   10296.7 &    7871.6 &   12776.1 \\ 
  0.0794 & 0.0912 &    6637.9 &    5106.3 &    8145.2 \\ 
  0.0912 & 0.1047 &    4521.1 &    3479.2 &    5503.3 \\ 
  0.1047 & 0.1202 &    3960.6 &    3167.7 &    4726.6 \\ 
  0.1202 & 0.1380 &    4326.7 &    3666.9 &    4971.0 \\ 
  0.1380 & 0.1585 &    1974.8 &    1460.5 &    2506.9 \\ 
  0.1585 & 0.1820 &    4019.1 &    2675.8 &    5461.1 \\ 
  0.1820 & 0.2089 &    4433.0 &    2889.0 &    6098.3 \\ 
  0.2089 & 0.2399 &    2357.5 &    1538.0 &    3251.3 \\ 
  0.2399 & 0.2754 &     673.6 &     130.6 &    1255.9 \\ 
  0.2754 & 0.6310 &     559.3 &     135.3 &     991.3 \\ 
  \hline
  \end{tabular}
\end{table}

In Table~\ref{tab:pk} we present the recovered marginalised power
spectrum band-powers calculated in 16 bins in $k$. These band-powers
were calculated by averaging over the MCMC chains described in
Section~\ref{sec:mcmc}, discounting the burn-in period for each. The
positions of the bins in $k$-space are also presented, as are
approximate errors for the points, calculated using the distribution
described in section~\ref{sec:like_fit}. In addition to these power
spectrum measurements, we find that $\beta(L_*,0)=0.55$ with the
1-$\sigma$ error region given by $0.46<\beta(L_*,0)<0.64$, again given
the distribution of $\beta(L_*,0)$ values described in the next
section.

The measured real-space power spectrum of $L_*$ galaxies is plotted in
Fig.~\ref{fig:real_pk_2dF}. This plot shows that, as expected,
although the band-powers extend beyond the $k$-range of modes selected
for analysis ($0.02<k<0.15\hompc$) because of the window function
caused by the survey geometry, they are only tightly constrained
within this range. The real-space power spectrum is shown to be well
matched to a standard concordance model $\Lambda$CDM power spectrum
with $\Omega_mh=0.2$, $\Omega_b/\Omega_m=0.15$,
$b(L_*,0)\sigma_8=0.8$, $n_s=1.0$ and $h=0.72$.

\section{comparison with previous galaxy power spectra}  
  \label{sec:pk_comparison}

\begin{figure}
  \setlength{\epsfxsize}{0.95\columnwidth} 
    \centerline{\epsfbox{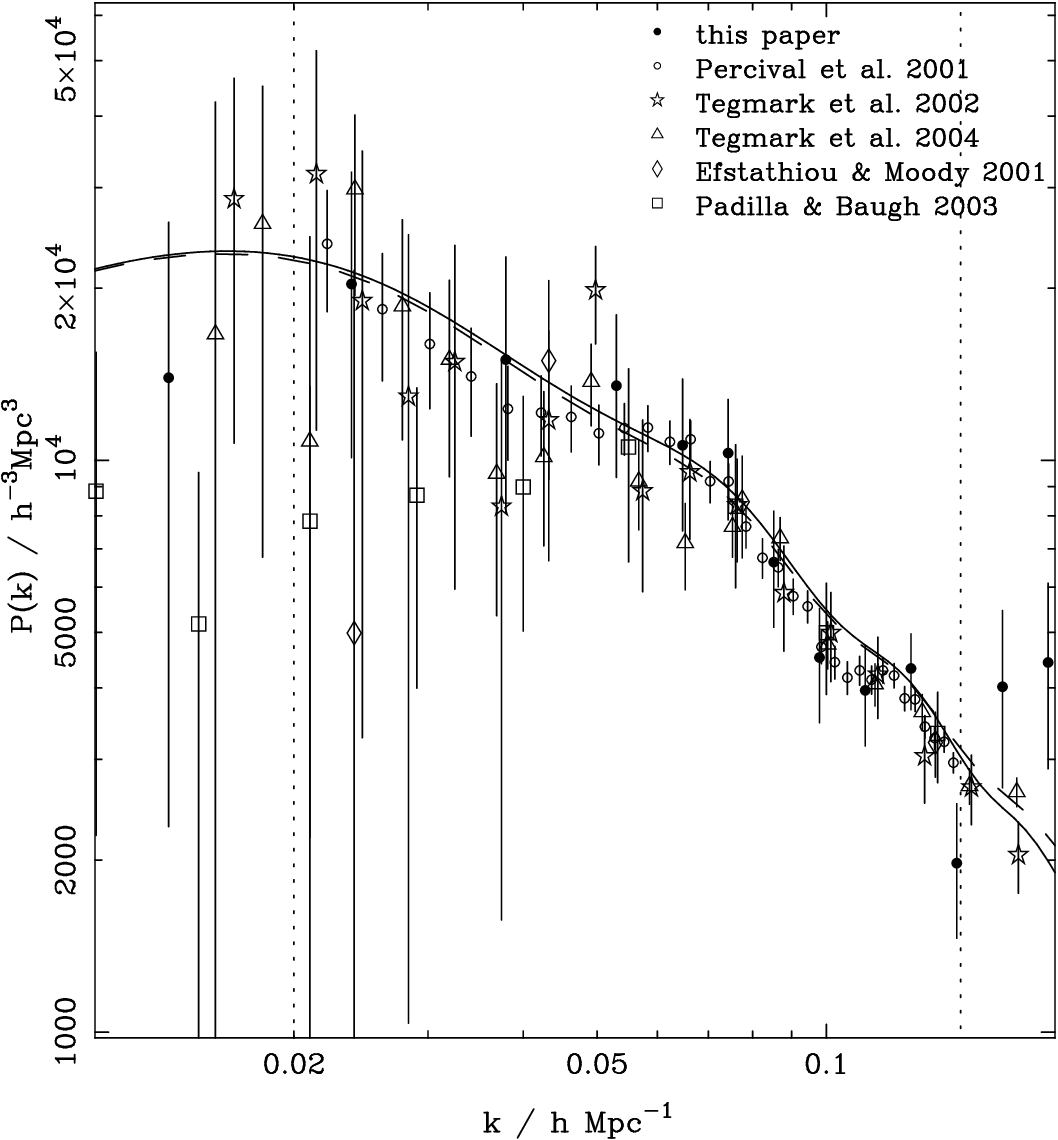}}

  \caption{Comparison of large-scale galaxy power spectra shapes
  calculated from various galaxy surveys using different analysis
  methods. Although errors are plotted for each measurement,
  comparison of different errors is complicated by window functions
  and correlations between different data sets. The power spectra data
  that cover a range in $k$-space have been approximately corrected
  for their window functions by multiplying by the net effect of the
  window on a model power spectrum with $\Omega_m.h=0.165$,
  $\Omega_b/\Omega_m=0.0$, $h=0.72$ \& $n_s=1$. For clarity, the
  $k$-ranges covered by different data points are not shown in this
  figure. The amplitude of each power spectra has been matched to the
  measured real-space power spectrum of 2dFGRS $L_*$ galaxies in this
  paper. The amplitude shifts required are given in
  Section~\ref{sec:pk_comparison}, where details of the different data
  sets can also be found. The relative normalizations of the different
  power spectra are potentially due to a combination of galaxy bias,
  redshift-space and evolutionary effects depending on the
  sample. Note that a normalization correction is even required for
  SDSS $L_*$ $P(k)$ data of Tegmark et~al. (2004) as, due to red
  selection, the SDSS $L_*$ galaxies are more biased than the 2dFGRS
  $L_*$ galaxies.  \label{fig:real_pk_cmpr}}
\end{figure}

The shape of the measured 2dFGRS real-space power spectrum of $L_*$
galaxies (solid circles) is compared with other measurements of galaxy
power spectra in Fig.~\ref{fig:real_pk_cmpr}. Because of correlations
between different data points in the same power spectrum and window
functions associated with some measurements, it is difficult to
compare error bars and the relative constraints on cosmological models
placed by different data sets. The easiest way to compare the
significance of the different data sets is to consider parameter
constraints for the same $k$-range and cosmological model. This is not
attempted in this paper. Instead, Fig.~\ref{fig:real_pk_cmpr} is
included simply to demonstrate the incredible consistency in the
large-scale ($0.02<k<0.15\hompc$) shape of recent measurements of
galaxy power spectra.

In Fig.~\ref{fig:real_pk_cmpr} we plot two estimates of the 2dFGRS
power spectrum shape in addition to the present analysis. The
redshift-space data of Percival et~al. (2001) are plotted as open
circles. These data were calculated from the $160000$ galaxy redshift
measured before February 2001, and $P(k)$ was multiplied by $0.69$ to
match the present data. This factor is expected to be due to a
combination of redshift-space effects and luminosity-dependent bias
(the Percival et~al. 2001 data are not corrected to correspond to
$L_*$ galaxies). We also plot the measurement of the 2dFGRS power
spectrum from the first $100000$ galaxies of measured by Tegmark,
Hamilton \& Xu (2002) -- open stars. These $P(k)$ data required
multiplication by a factor $0.61$ to match the amplitude of the
real-space power spectrum measured in this work. The Tegmark
et~al. (2002) analysis method was strongly affected by
luminosity-dependent bias, potentially leading to a tilt in the
recovered power spectrum, and a different overall amplitude (Percival,
Verde \& Peacock 2003).

We also plot the SDSS power spectrum of Tegmark et~al. (2004) as open
triangles in Fig.~\ref{fig:real_pk_cmpr}. The data of Tegmark
et~al. (2004) are crudely corrected for bias by dividing by an
effective (bias)$^2$ at each $k$-value. While this corrects for an
overall normalization offset, there is an additional minor effect
caused by the change in survey geometry that is not corrected in this
approach (Percival, Verde \& Peacock 2003). In the Spherical Harmonics
analysis of the 2dFGRS data presented here, the data are corrected for
variable bias in a self-consistent way. The SDSS data of Tegmark
et~al. (2004) required a downward correction by a factor $0.77$ to
match the normalization of the 2dFGRS data, presumably because of a
difference in the bias of 2dFGRS $L_*$ and SDSS $L_*$ galaxies.

We have also compared our power spectrum to that recovered by
deprojecting the APM survey, the parent survey to the 2dFGRS. Results
from analyses of Efstathiou \& Moody (2001) -- open diamonds, and
Padilla \& Baugh (2003) -- open squares, are plotted and shown to be
in satisfactory agreement. There is weak evidence that these power
spectra have less power on large scales ($k\sim0.03$) than the data
from the redshift surveys, but given the large errors on the
deprojected power spectra, this deviation is not significant. The
offsets required for these data sets to match the 2dFGRS $P(k)$
normalization were $0.79$ and $0.62$ for the data of Efstathiou \&
Moody (2001) and Padilla \& Baugh (2003) respectively.

\section{fitting to the likelihood distribution}  \label{sec:like_fit}

In this Section we use the MCMC chains described in
Section~\ref{sec:mcmc} to determine a simple model of the likelihood
surface. If, to the accuracy of the present power spectrum
constraints, there is little data loss compressing the model power
spectrum into the 16 band-powers described in
Section~\ref{sec:bands_choice}, then this will provide a method for
quickly estimating the likelihood of a cosmological model.

The basic problem is to find an accurate description of the
$17$-parameter likelihood hyper-surface. From Eq.~\ref{eq:like_allsky}
it is immediately clear that, even for an all-sky volume-limited
noiseless survey of a clustered field, the true likelihood is not
symmetric in $P^{\rm TH}(k_i)$ about any point. This rules out a
Gaussian likelihood distribution of the form
\be
  -2\ln{\cal L} \propto \sum_{ij}[Q_i^{\rm TH}-Q_i]
      M_{ij}[Q_j^{\rm TH}-Q_j],
  \label{eq:like_gauss}
\ee
with $Q_i=P(k_i)$ for $1<i<16$, $Q_{17}=\beta(L_*,0)$ and fixed
inverse covariance matrix $M_{ij}$. Assuming such a distribution would
result in a biased estimate of the likelihood surface. 

This could be corrected either by allowing the inverse covariance
matrix to be a function of the model being tested or by considering a
different form for the likelihood, for example by allowing $Q_i =
f_i[P(k_i)]$, $1<i<16$ and $Q_{17}=f_{17}[\beta(L_*,0)]$ to be more
general functions of the band powers and $\beta(L_*,0)$. In this case,
$M_{ij}$ would be assumed to be independent of the model to be tested.

This problem is analogous to the compression of CMB maps into a series
of band-powers and has been considered in that context by a number of
authors. In particular, Bond et~al. (2000) suggest assuming an offset
lognormal likelihood distribution for CMB band powers. For the
real-space band powers and $\beta(L_*,0)$ this corresponds to assuming
that the likelihood is given by Eq.~\ref{eq:like_gauss}, but with $Q_i
= \ln[P(k_i)]\bar{P}_i$, $1<i<16$ where $\bar{P}_i$ is the maximum
likelihood band-power, and similarly for $\beta(L_*,0)$. Bond
et~al. (2000) argued that this change of variables retains the true
curvature matrix calculated from Eq.~\ref{eq:like_true} in the case of
an all-sky CMB map with no noise (i.e. the correct behaviour of the
likelihood around the peak). In addition, for an all-sky CMB map with
no noise, the offset log-normal distribution is approximately skewed
beyond the peak in the expected fashion given the dependence on the
theoretical power displayed by the true likelihood. This argument also
applies to the galaxy distribution being analysed in this paper.

Although both of the Gaussian and offset log-normal distributions
provide the correct curvature around the peak of the likelihood, it is
not clear that either provides an unbiased likelihood surface beyond
this point, particular when noise and correlations become
important. In a recent analysis of the WMAP CMB data, Verde
et~al. (2003) considered a combination of Gaussian and lognormal
likelihoods, which automatically matches the peak curvature, but
additionally matches the behaviour around the peak for the case of an
all-sky CMB map with no noise. In order to do this, a likelihood
distribution of the form $\ln{\cal L} = 1/3\ln{\cal L}_{\rm
Gauss}+2/3\ln{\cal L}_{\rm OLN}$ was adopted, where ${\cal L}_{\rm
Gauss}$ and ${\cal L}_{\rm OLN}$ are the Gaussian and offset lognormal
likelihoods respectively.

\begin{table}
  \caption{Parameters of the likelihood surface fits given by
  Eqns.~\ref{eq:like_gauss} \&~\ref{eq:like_special}. For each
  band-power these data were calculated by fitting the MCMC output
  marginalised over the remaining $16$ parameters, as described in
  Section~\ref{sec:like_fit}. In addition $a=0.689$ and
  $\bar{\beta}(L_*,0)=0.551$ were obtained from the fit to the
  marginalised likelihood as a function of
  $\beta(L_*,0)$. \label{tab:like_fit}}

  \centering \begin{tabular}{cccc} \hline
  \multicolumn{2}{c}{$k$ / $\hompc$} & $a$ & $\bar{P}(k)$ / $\mpcohcube$ \\ 
  min & max &  & \\
  \hline
  0.0100 & 0.0174 &     0.243 &    9192.4 \\ 
  0.0174 & 0.0302 &     0.543 &   19588.3 \\ 
  0.0302 & 0.0457 &     0.542 &   14547.5 \\ 
  0.0457 & 0.0603 &     0.544 &   13326.2 \\ 
  0.0603 & 0.0692 &     0.539 &   10476.3 \\ 
  0.0692 & 0.0794 &     0.564 &   10194.9 \\ 
  0.0794 & 0.0912 &     0.483 &    6534.5 \\ 
  0.0912 & 0.1047 &     0.465 &    4429.4 \\ 
  0.1047 & 0.1202 &     0.486 &    3907.2 \\ 
  0.1202 & 0.1380 &     0.473 &    4292.9 \\ 
  0.1380 & 0.1585 &     0.503 &    1948.7 \\ 
  0.1585 & 0.1820 &     0.914 &    4046.9 \\ 
  0.1820 & 0.2089 &     1.000 &    4493.6 \\ 
  0.2089 & 0.2399 &     0.770 &    2357.5 \\ 
  0.2399 & 0.2754 &     0.514 &     542.9 \\ 
  0.2754 & 0.6310 &     0.557 &     470.0 \\ 
  \hline
  \end{tabular}
\end{table}

\begin{figure*}
  \setlength{\epsfxsize}{0.9\textwidth} 
    \centerline{\epsfbox{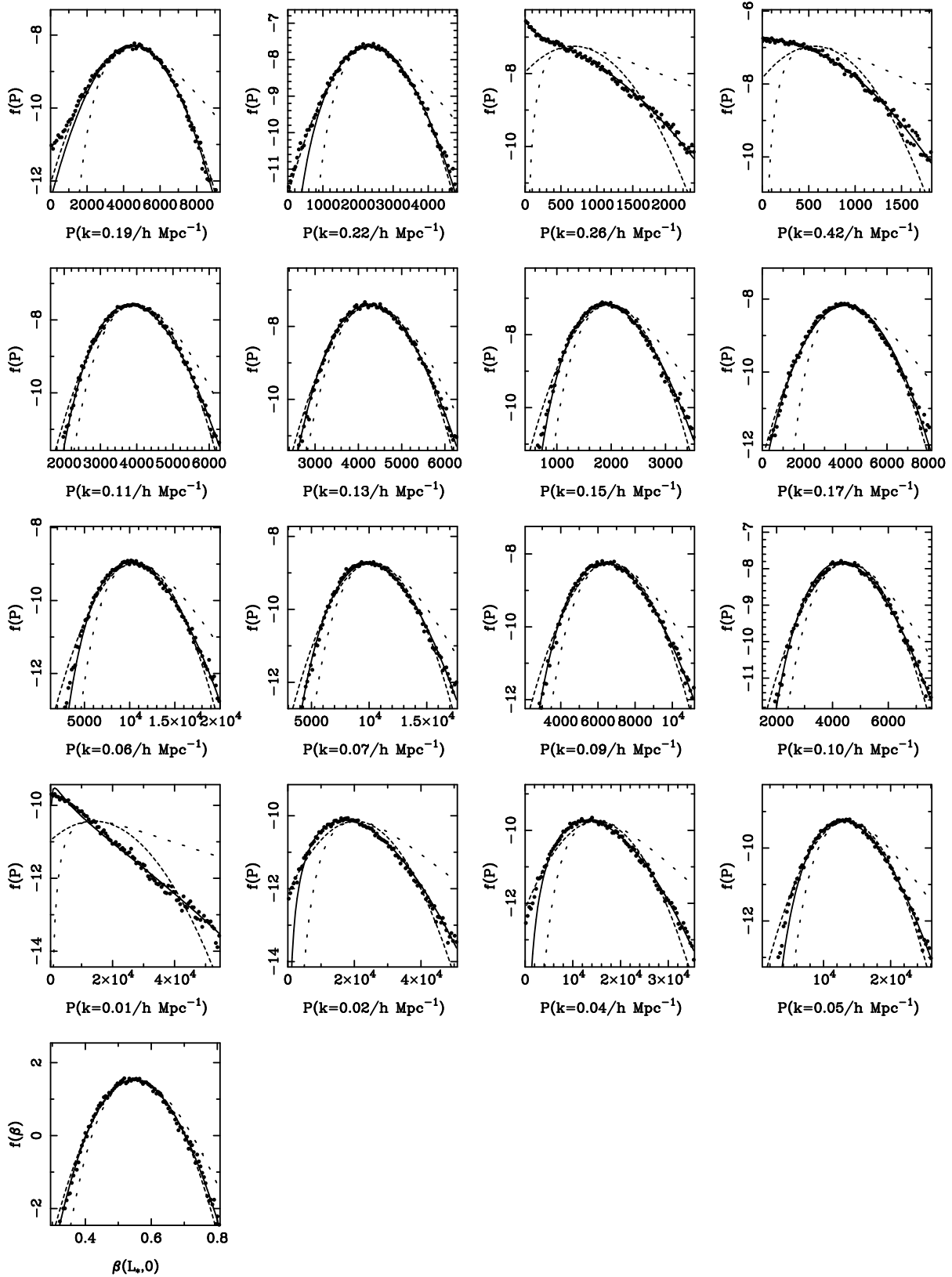}}

  \caption{The distribution of the 16 recovered band-power values and
  $\beta(L_*,0)$, calculated from the MCMC chains. Each distribution
  has been calculated after marginalising over the remaining 16
  parameters (solid symbols). This distribution has been fitted with a
  Gaussian (dashed line), and a log-normal distribution with maximum
  at the distribution mean (dotted line). We have also fitted the
  distribution assuming that the distribution is Gaussian, not in the
  band-power (or $\beta(L_*,0)$), but in a linear combination of the
  band-power and its logarithm (solid line). \label{fig:band_fit}}
\end{figure*}

However, for the Spherical Harmonics analysis considered here, both
shot noise and coupling between modes are significant, and the
behaviour around the likelihood peak is not so easily estimated
analytically. The good news is that the MCMC technique produces a
sampling of the likelihood surface that can be fitted. Given a chain
of values that sample the hyper-surface, estimating the likelihood
distribution for a given parameter, marginalising over the other 16
parameters simply involves binning all of the values in the chain. For
the band-powers and $\beta(L_*,0)$ the marginalised distributions were
found to be well fitted by Gaussian distributions in
\be
  Q_i = a_iP_i + (1-a_i)\ln[P_i]\bar{P}_i,
    \label{eq:like_special}
\ee
the form of which was motivated by the Gaussian and offset lognormal
distributions discussed above. Here, we define $P_i\equiv P(k_i)$ for
$1<i<16$ and $P_{17}\equiv\beta(L_*,0)$ and similarly for the maximum
likelihood positions $\bar{P}_i$. The best fit values of $\bar{P}_i$
and $a_i$ (with the constraint $0<a_i<1$) were fitted for each
band-power and $\beta(L_*,0)$ to the distribution of parameter values
in the chains, and are given in Table~\ref{tab:like_fit}. Note that
the maximum likelihood band-powers $\bar{P}(k_i)$ differ from the
marginalised band-powers given in Table~\ref{tab:pk} as expected if
the distributions are not symmetric. Plots showing the fits are
presented in Fig.~\ref{fig:band_fit}. Outside of the range of nodes
probed in P04 ($0.02<k<0.15\hompc$) the probability distribution of
band-powers is strongly non-Gaussian. However, assuming Gaussianity in
$Q_i$ (Eq.~\ref{eq:like_special}) rather than the band-powers provides
a reasonable fit to the recovered distribution. The covariance matrix
was found from the MCMC chains assuming Gaussianity in $Q_i$
(i.e. from the expected multiples $\langle Q_i Q_j \rangle$), and was
inverted to give $M_{ij}$. The best-fit values of $\bar{P}_i$, $a_i$
and the corresponding inverse covariance matrix $M_{ij}$ are available
from {\tt http://www.roe.ac.uk/{\tt\char'176}wjp/}.

\section{testing the likelihood surface} \label{sec:simple_models}

\begin{figure}
  \setlength{\epsfxsize}{0.9\columnwidth} 
    \centerline{\epsfbox{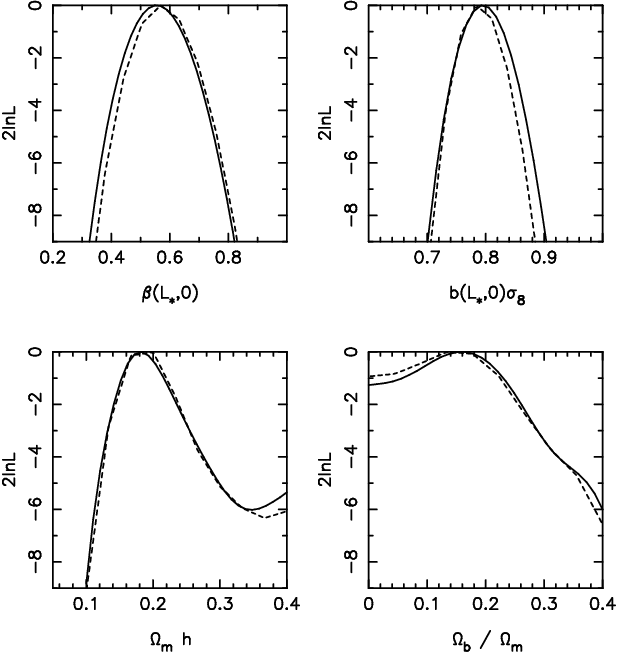}}

  \caption{Marginalised likelihoods for the 4 parameter $\Lambda$CDM
  cosmological model considered by P04 with $h=0.72$ and
  $n_s=1.0$. The dashed line was calculated directly from the
  Spherical Harmonics procedure -- for each model, a covariance matrix
  is constructed and applied to calculate the likelihood. The solid
  lines were calculated assuming that the 16 $P(k)$ measurements and
  $\beta(L_*,0)$ have a Gaussian distribution in the functions given
  by Eq.~\ref{eq:like_special}. \label{fig:like_models_1d}}
\end{figure}

\begin{figure}
  \setlength{\epsfxsize}{0.9\columnwidth} 
    \centerline{\epsfbox{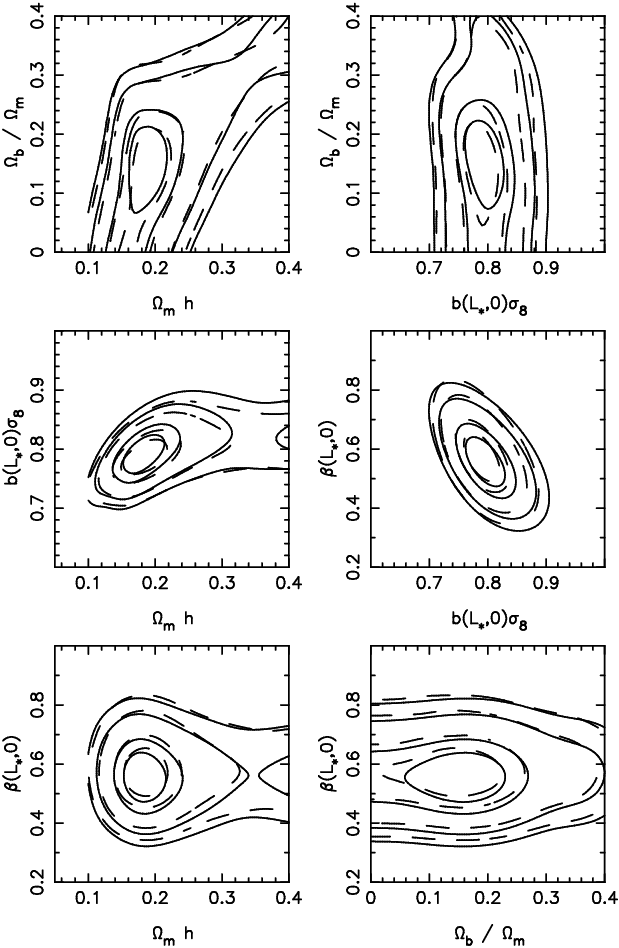}}

  \caption{Contour plots showing changes in the likelihood from the
  maximum of $2\Delta\ln{\cal L}=1.0, 2.3, 6.0, 9.2$ for different
  parameter combinations for the combined likelihood from the 2dFGRS
  NGP and SGP catalogues, assuming a $\Lambda$CDM power spectrum with
  $h=0.72$ and $n_s=1.0$. Dashed contours were calculated directly
  from the Spherical Harmonics procedure and correspond to those in
  fig.~6 in P04. Solid contours were calculated assuming that the 16
  $P(k)$ measurements and $\beta(L_*,0)$ have a Gaussian distribution
  in the functions given by Eq.~\ref{eq:like_special}. Solid and
  dashed lines therefore correspond to the 1D marginalised likelihood
  curves presented in Fig.~\ref{fig:like_models_1d}.
  \label{fig:like_models_2d}}
\end{figure}

The 4-parameter set of cosmological models considered by P04 has been
used to compare the estimated likelihoods with those calculated
directly from the Spherical Harmonics procedure, where for each model
a covariance matrix is constructed and used to calculate the
likelihood through Eq.~\ref{eq:like_true}. In this set of models, the
matter density multiplied by the Hubble constant $\Omega_mh$, baryon
fraction $\Omega_b/\Omega_m$, power spectrum normalization
$b(L_*,0)\sigma_8$ and $\beta(L_*,0)$ were allowed to vary, while
scalar spectral index $n_s=1.0$ \& Hubble constant $h=0.72$ were
fixed. Model band-powers were calculated by averaging the Eisenstein
\& Hu (1998) power spectrum fits over each bin in $\log(k)$. The power
spectra were extrapolated into the non-linear regime using the fitting
formulae of Smith et~al. (2003).

Fig.~\ref{fig:like_models_1d} compares the 1d marginalised likelihoods
from the two methods of likelihood calculation. Good agreement is
demonstrated between the true and estimated likelihood surfaces. There
is weak evidence for a slight bias to slightly higher
$b(L_*,0)\sigma_8$ and lower $\beta(L_*,0)$ values. In
Fig.~\ref{fig:like_models_2d} we show contour plots of the
marginalised likelihood for each 2-parameter combination (our 4
parameter model has 6 distinct pairs of parameters). This shows that
$b(L_*,0)\sigma_8$ and $\beta(L_*,0)$ are anti-correlated and the
offsets in $b(L_*,0)\sigma_8$ and $\beta(L_*,0)$ are therefore
related. The difference could be caused by inaccuracies in the
likelihood model fits of Eq.~\ref{eq:like_special}, or due to
information loss because of the data compression into $17$
parameters. However, the offset is significantly smaller than any
errors that would be placed on either parameter from this analysis,
and is not therefore of major concern.

\section{summary \& discussion}

A new method using the MCMC technique has been developed to recover the
real-space power spectrum from a decomposition of a galaxy overdensity
field and has been applied to the 2dFGRS Spherical Harmonics analysis
presented in P04. The recovered 2dFGRS $L_*$ power spectrum was
presented as $16$ marginalised power spectrum band-powers. The MCMC
technique was also used to obtain information about the shape of the
likelihood surface. Using this information, the likelihood surface was
fitted with a simple model, providing a quick method for estimating
the likelihood of different cosmological models.

The advantage of the MCMC method over previous deconvolution routines
lies in providing a map of the likelihood surface. Maximum likelihood
techniques have previously been used to recover the real-space power
spectrum in a series of band-powers (Ballinger, Heavens \& Taylor
1995), using a standard routine to iterate to the likelihood
maxima. However, such a step-wise maximum likelihood method only
partially explores the likelihood surface giving poorly determined
conditional errors (Ballinger et~al. 1995; Tadros et~al. 1999). The
MCMC technique provides a significant advance over such methods, as it
enables a more complete exploration of the likelihood surface to be
quickly undertaken.

An alternative to the step-wise approach to determining ML
band-powers, is to use quadratic band-power estimators (Hamilton
1997a;b) favoured by Tegmark et~al. (2002; 2004). The band-powers that
we model have no window functions, and therefore must correspond to
the linear combinations of the quadratic estimators given by the
inverse of the Fisher information matrix (see, for example, section
3.5 of Tegmark et~al. 2002). In this approach, the resulting
band-powers are assumed to have a Gaussian likelihood with
model-independent covariances, which, as we argued in
Section~\ref{sec:like_true}, for an all-sky noiseless survey must lead
to a different form for the likelihood compared with the true
likelihood given by Eq.~\ref{eq:like_allsky}. This is consistent with
the MCMC analysis presented in this paper which suggests that the
likelihood distribution around the band-powers is not well described
by a multi-variate Gaussian distribution. Either the covariance matrix
must change its form, or equivalently the shape of the function
assumed for the likelihood surface must change. The MCMC technique is
therefore advantageous over the quadratic band-power method as it
provides the information required to allow the likelihood surface to
be accurately modelled beyond the likelihood peak.

Another factor that makes the MCMC routine particularly well suited to
this application is the nature of the likelihood surface. No evidence
was found for multiple likelihood maxima, which would slow the MCMC
mapping. Additionally, more simple techniques would fail because the
surface does not follow an obvious form -- for instance it is close
to, but not exactly a multi-variate Gaussian so a simple fit of a
multi-variate Gaussian would be inappropriate. Fitting to the
likelihood surface has been shown to provide an accurate estimate of
the combined likelihoods for a number of cosmological models covering
a range of power spectrum shapes. This gives us confidence that it can
be extended to cover a wider range of cosmological models.

From the MCMC chains, we have estimated marginalised band-powers, and
have compared these with measurements of the real-space power spectrum
from the SDSS, and with other measurements from the 2dFGRS. Excellent
agreement is observed and the basic shape of the power spectrum now
seems to be well constrained. We have not directly compared with the
real-space power spectrum of the IRAS 1.2Jy galaxies (Ballinger
et~al. 1995) or with the real-space power spectrum of the PSCz
galaxies estimated by Tadros et~al. (1999), in order not to further
clutter Fig.~\ref{fig:real_pk_cmpr}. However, it is worth noting that
the general shape of the recovered power spectra are similar. We have
compared with the deprojected APM data of Efstathiou \& Moody (2001)
and Padilla \& Baugh (2003), although the 2dFGRS and SDSS data
represent a significant improvement over these data.

In addition to the band-power measurements we simultaneously fit the
linear redshift-space distortions. A best-fit value of
$\beta(L_*,0)=0.55$ was recovered with 1-$\sigma$ confidence region
given by $0.46<\beta(L_*,0)<0.64$. This compares very well with the
value of $\beta(L_*,0)=0.58\pm0.08$ recovered in P04 marginalising
over the basic set of cosmological models considered. Note that from
the approximation to the likelihood we recover the same parameter
value if we consider the range of models chosen, and their associated
priors on the shape of the power spectrum.

The marginalised band-power and $\beta(L_*,0)$ measurements were
presented in Table~\ref{tab:pk}. These should not be confused with the
parameters of the fit to the likelihood surface that were presented in
Table~\ref{tab:like_fit} and associated covariance matrix. All of
these data are publically available from {\tt
http://www.roe.ac.uk/{\tt\char'176}wjp/}, where simple segments of
code demonstrating how the likelihood can be quickly calculated are
also provided. It should therefore be possible to easily combine the
cosmological constraints provided by the Spherical Harmonics analysis
of P04 with other cosmological probes.

Obviously, these constraints should not be combined with other
analyses of the 2dFGRS such as provided by Percival et~al. (2001) or
Cole et~al. (2004) as these results are strongly inter-dependent. The
analysis presented by Cole et~al. (2004) uses more galaxies and
therefore provides tighter constraints on the power spectrum shape
than the Spherical Harmonics analysis considered here and should be
used in preference to this work if the power spectrum shape is the
important consideration. The advantage of the Spherical Harmonics
analysis is that, although it provides weaker information on $P(k)$
shape, it provides additional correlated information about
$\Omega_m^{0.6}\sigma_8$ from fitting the redshift-space distortions.

An example application where using Spherical Harmonics rather than
Fourier likelihoods may be advantageous is to provide constraints on
the neutrino mass density. Here, the amplitude of the low-redshift
matter power spectrum provides a significant constraint, and has
previously been determined by converting a redshift-space power
spectrum amplitude to $\sigma_8$ using $\beta$ and galaxy bias $b$
calculated from other analyses (e.g. Spergel et al. 2003). The problem
with such an approach is that $\beta$ and $b$ are often measured on
different scales, using different analysis techniques and model
dependencies. The Spherical Harmonics method instead sets a constraint
on $\sigma_8$ (through $\Omega_m^{0.6}\sigma_8$) in a self-consistent
way. Combining the Spherical Harmonics likelihoods with CMB
likelihoods should therefore provide a more robust determination of
the neutrino mass density.

\section*{ACKNOWLEDGMENTS}

The author wishes to thank the staff of the Anglo-Australian
Observatory and the 2dFGRS team members whose dedicated work provided
this wonderful survey data. Useful discussions with Alan Heavens and
John Peacock concerning likelihood calculation and the Spherical
Harmonics method were also of great help. The anonymous referee's
comments led to a significant improvement in the presentation of this
work. WJP is supported by PPARC through a Postdoctoral Fellowship.

\end{document}